# Classical LQG's limitation and modification in stochastic vibration system

Yinmiao Luo

*(Department of Mechanics, School of Aeronautics and Astronautics, Zhejiang University, Hangzhou 310027, China)*

**Abstract**:The function of control force is deduced by stochastic averaging method in shochastic vibration system.It is found that classical LQG is not full optimization because control force from displacement is of no effect to depress stochastic response.A modified vLQG control strategy is proposed . Numerical result shows that vLQG has better control effectiveness comparing with classical LQG control strategy.

**Keywords**:Stochastic averaging,LQG control,Stochastic Vibration,Hamilton

## 1. Introduction

Linear-quadratic-Gaussian (LQG) control strategy is widely used in dynamic system(Anderson and Moore 1990). LQG was deduced from Linear-quadratic (LQ) state-feedback regulator for state-space system,and accounted for model experienced disturbances(Colaneri and de Nicolao 1995, Chen and Dong 1989). When it is application to stochastic vibration system,people consider that LQG is optimization in state space.Because of it's applicability and being implemented in practical applications,LQG is mostly used in stochastic vibration control(Iourtchenko 2009).If load applied to vibration system is stochastic,for example,white noise or wideband noise,it is found that LQG is not full optimality.It's control efficiency can be improved.

## 2. Stochastic averaging methods for quasi-integrable Hamiltonian system

Most mechanical vibration system is quasi-integrable Hamiltonian systems .Considering a controlled strongly nonlinear conservative oscillator subject to lightly linear or nonlinear damping and weakly excitation of wide-band random processes, The motion equation of the system is of the form

$$\frac{d^2 X}{dt^2} + g(X) = \varepsilon c(X,\dot{X}) + \varepsilon u(X,\dot{X}) + \varepsilon^{\frac{1}{2}} f \xi(t) \qquad (1)$$

where g (X) represents elastic restoring force which could be strongly non-linear, ε is a small positive parameter, $\varepsilon c(X,\dot{X})$ represents light linear or non-linear damping; $\varepsilon u(X,\dot{X})$ represents weak feedback control force; $f\xi_k(t)$ are wide-band stationary random processes with zero mean and it's spectral densities S(ω).

The Hamilton function which can be expressed as total energy of the vibration system is

$$H = \frac{1}{2}\dot{X}^2 + V(X)$$
$$V(X) = \int_0^X g(s)ds \qquad (2)$$

Where V(X) is system potential energy.Assume that when ε = 0, the Hamiltonian system with

Hamilton function H has a trivial solution $X=B$, $\dot{X}=0$ and a family of periodic solution around the trivial solution. When ε is small, system (1) has periodic stochastic solutions around the trivial solution. The system governed by equation (1) without stochastic excitation has been studied by Xu and Cheung(Xu and Chung 1994). The sample solution of system (1) with stochastic excitation can be assumed of the following form(Huang and Zhu 2004, Huang and Zhu 2002, Huang and Zhu 1997)

$$X(t) = A\cos\Theta(t) + B$$
$$\dot{X}(t) = -A\nu(A,\Theta)\sin\Theta(t) \quad (3)$$
$$\Theta(t) = \Phi(t) + \Gamma(t)$$

where

$$\nu(A,\Theta) = \frac{d\Phi}{dt} = \frac{\sqrt{2[V(A+B) - V(A\cos\Theta + B)]}}{|A\sin\Theta|} \quad (4)$$

and A, B, Φ, Ψ and Θ are all random processes. Taking Eqs.(3) as a generalized van der Pol transformation, the following equations for A and Θ can be obtained:

$$\frac{dA}{dt} = \varepsilon F_1(A,\Gamma) + \varepsilon F_1^u(A,\Gamma,u) + \varepsilon^{1/2}G_1(A,\Gamma)\xi(t),$$
$$\frac{d\Gamma}{dt} = \varepsilon F_2(A,\Gamma) + \varepsilon F_2^u(A,\Gamma,u) + \varepsilon^{1/2}G_2(A,\Gamma)\xi(t) \quad (5)$$

where

$$\varepsilon F_1(A,\Gamma) = \frac{-A}{g(A+B)(1+h)} c(A\cos\Theta + B, -A\nu(A,\Theta)\sin\Theta) \times \nu(A,\Theta)\sin\Theta$$

$$\varepsilon F_1^u(A,\Gamma) = \frac{-A}{g(A+B)(1+h)} u \times \nu(A,\Theta)\sin\Theta$$

$$\varepsilon F_2(A,\Gamma) = \frac{-1}{g(A+B)(1+h)} c(A\cos\Theta + B, -A\nu(A,\Theta)\sin\Theta) \times \nu(A,\Theta)\cos\Theta$$

$$\varepsilon F_2^u(A,\Gamma) = \frac{-1}{g(A+B)(1+h)} u \times \nu(A,\Theta)\cos\Theta \quad (6)$$

$$\varepsilon^{1/2}G_1(A,\Gamma) = \frac{-A}{g(A+B)(1+h)} f \times \nu(A,\Theta)\sin\Theta$$

$$\varepsilon^{1/2}G_2(A,\Gamma) = \frac{-1}{g(A+B)(1+h)} f \times \nu(A,\Theta)\cos\Theta$$

The explicit expression for *h* is

$$h = \frac{dB}{dA} = \frac{g(-A+B) + g(A+B)}{g(-A+B) - g(A+B)} \quad (7)$$

Equation (5) can be modelled as Stratonovich stochastic differential equation and then transformed it into Ito stochastic differential equation(here Wong Zakai item equals to zero).

$$dA = [m(A) + \varepsilon\langle F_1^u(A,\Gamma)\rangle_t]dt + \sigma(A)dB(t) \quad (8)$$

where *B* (*t*) are unit Wiener processes. The drift and diffusion coeffcients in Ito equation (8) are functions of slowly varying processes *A* and *Γ* and rapidly varying process *Θ*. Averaging them with respect to *Θ* yields the following averaged Ito equations:

$$m(A) = \varepsilon \left\langle F_1 + \int_{-\infty}^{0} \left( \left. \frac{\partial G_{1k}}{\partial A} \right|_t G_{1l}|_{t+\tau} + \left. \frac{\partial G_{1k}}{\partial \Gamma} \right|_t G_{2l}|_{t+\tau} \right) R_{kl}(\tau) d\tau \right\rangle_\Theta \quad (9)$$

$$\sigma^2(A) = \varepsilon \left\langle \int_{-\infty}^{\infty} G_{1k}|_t G_{1l}|_{t+\tau} R_{kl}(\tau) d\tau \right\rangle_\Theta$$

Where $\langle \cdot \rangle_\Theta$ means averaging over $\Theta$.

The partially averaged Itˆo equation for the energy of the vibration system can be obtained from Eq.(9) using Ito differential rule

$$dH = [\bar{m}_1(H) + \bar{m}_2(H,u)]dt + \bar{\sigma}(H)dB(t) \quad (10)$$

Where

$$\bar{m}_1(H) = \left[ g(A+B)(1+h)m(A) \right]\Big|_{A=V^{-1}(H)-B}$$

$$\bar{m}_2(H,u) = \left[ g(A+B)(1+h)\varepsilon \left\langle F_1^u(A,\Gamma) \right\rangle \right]\Big|_{A=V^{-1}(H)-B} \quad (11)$$

$$\bar{\sigma}^2(H) = \left[ g^2(A+B)(1+h)^2 \sigma^2(A) \right]\Big|_{A=V^{-1}(H)-B}$$

The FPK equation associated with Ito equation (10) is of the form

$$\frac{\partial p}{\partial t} = -\frac{\partial}{\partial A}\left\{\left[\bar{m}_1(H) + \bar{m}_2(H,u)\right]p\right\} + \frac{1}{2}\frac{\partial^2}{\partial H^2}\left[\bar{\sigma}^2(H)p\right] \quad (12)$$

The exact stationary solution to FPK Eq. (12) can be obtained as follows:

$$p(H) = \frac{C}{\bar{\sigma}^2(H)} \exp\left[ \int_0^H 2\frac{\bar{m}_1(\chi) + \bar{m}_2(x,u(\chi))}{\bar{\sigma}^2(\chi)} d\chi \right] \quad (13)$$

$$p_c(\dot{x}, x) = \left. p_c(H) \middle/ T(H) \right|_{H=H(\dot{x},x)}$$

$$E(H_c^2) = \int_0^\infty H^2 p_c(H) dH$$

$$E(H_u^2) = \int_0^\infty H^2 p_u(H) dH \quad (14)$$

$$E(u^2) = \int_0^\infty u^2 p_c(H) dH$$

$$E(u_L^2) = \int_0^\infty u_L^2 p_L(\dot{x},x) d\dot{x}dx$$

## 3. The limitation of classical LQG strategy in stochastic vibration control

If damping c is linear damping and g(x) is linear in Eq(1), then $B=0, h=0$, The control force can be designed with LQG control strategy as

$$u(t)|_{LQG} = [G(1) \quad G(2)][X(t) \quad \dot{X}(t)]^T \quad (15)$$

Because LQG control strategy is feedback control based on state space, the regulator coefficient is negative for single degree of freedom or in mode space. After stochastic averaging, drift coefficient about control force can be deduced

$$\bar{m}_2(H,u)\big|_{LQG} = G(2)H \tag{16}$$

This means that control effect from (13) is independent of G(1),which is about displacement feedback.If we define a velocity feedback from LQG as vLQG:

$$u(t)\big|_{vLQG} = G(2)\dot{X}(t) \tag{17}$$

The mean square displacement control force of LQG is

$$\begin{aligned} E\left(u^2\big|_{LQG}\right) &= G^2(2)E\left(\dot{x}^2\right) + G^2(1)E\left(x^2\right) + 2G(1)G(2)E(x\dot{x}) \\ &= E\left(u^2\big|_{vLQF}\right) + G^2(1)E\left(x^2\right) + 2G(1)G(2)E(x\dot{x}) > E\left(u^2\big|_{vLQG}\right) \end{aligned} \tag{18}$$

So the control force of LQG is larger than vLQG with same control effectiveness.This means vLQG is much more optimized and has much efficiency.

## 4. Example

4.1 Control of SDOF linear system excited by external excitation

As an example to illustrate the advantage of vLQG, firstly consider the following controlled single degree of freedom linear system excited by external excitation of wide-band stationary random processes:

$$\frac{d^2X}{dt^2} + \zeta\dot{X} + \omega_0^2 X = u(X,\dot{X}) + f\xi(t) \tag{19}$$

Where $\zeta = 0.02, \omega_0 = 1$. $u$ is a feedback control force; $f\xi_1$ is produced from transfer function

$$G_t(s) = \frac{G_0\omega_n^2}{s^2 + 2\zeta_n\omega_n s + \omega_n^2} \tag{20}$$

It is stationary with zero mean and spectral densities

$$S_f(\omega) = \frac{G_0\omega_n^2}{\sqrt{\left(\omega_n^2 - \omega^2\right)^2 + \left(2\zeta_n\omega_n\omega\right)^2}} \tag{21}$$

$$G_0 = 1, \zeta_n = 0.1, \omega_n = 2$$

The control results are listed in Table 1. It is seen from Tables 1 that vLQG control strategy is more effective than LQG control strategy.

Table 1    Numerical results of LQG and vLQG for SDOF: standard deviation

| Control strategy | Without control | LQG | vLQG |
| --- | --- | --- | --- |
| Response H | 0.4453 | 0.00624 | 0.005384 |
| Control Force u | / | 0.05504 | 0.05498 |

## 4.2 Control of MDOF linear system excited by external excitation

vLQG is also available to multi-degree-of-freedom (MDOF) linear system excited by external excitation. Consider a controlled linear frame structure under support excitations such as earthquake. Its equation of motion can be expressed as

$$M\ddot{X} + C\dot{X} + KX = -ME\ddot{X}_g + PU \qquad (22)$$

where $X = [x_1 \ x_2 \ x_3]^T$, $X_i$ is the horizontal displacement of $i$th floor relative to ground; $M$, $C$, $K$ are the $n \times n$ mass, damping and stiffness matrices, respectively; $E = [1 \ 1 \ 1]^T$; $\ddot{X}_g$ is the horizontal ground acceleration; $U$ is the control force produced by control device located at top floor; $P$ is the $n \times 1$ control device placement matrix.

Introduce the following modal transformation

$$X = \Phi q \qquad (23)$$

where $\Phi$ is the $n \times n$ real modal matrix of the structure. Eq. (1) becomes

$$\ddot{q} + 2\zeta\Omega\dot{q} + \Omega^2 q = -\beta\ddot{X}_g + u \qquad (24)$$

where $q = [q_1 \ q_2 \ q_3]^T$, $Q_i$ is the generalized displacement of $i$th mode; $\Omega^2 = diag[\omega_i^2] = \Phi^T K \Phi$, $\omega_i$ is the natural frequency of $i$th mode; $2\zeta\Omega = \Phi^T C \Phi$, $\zeta = diag[\zeta_i]$ $\zeta_i$ is the damping ratio of $i$th mode; $\beta = [\beta_1 \ \beta_2 \ \beta_3]^T = \Phi^T ME$; $u = [u_1 \ u_2 \ u_3]^T = \Phi^T PU$ and $u_i$ is the generalized control force of $i$th mode. Coefficient matrixs corresponding (22)~(24) are

$$\Omega = 2\pi \begin{bmatrix} 1.583 & & \\ & 4.436 & \\ & & 6.412 \end{bmatrix}, \zeta = \begin{bmatrix} 0.9951 & & \\ & 2.7883 & \\ & & 5.0244 \end{bmatrix},$$

$$K = \begin{bmatrix} 99 & & \\ & 777 & \\ & & 1623 \end{bmatrix}, \Phi = \begin{bmatrix} -0.0005 & 0.0012 & -0.0009 \\ -0.0009 & 0.0005 & 0.0012 \\ -0.0012 & -0.0009 & -0.0005 \end{bmatrix}$$

$G_0 = 0.16, \zeta_n = 0.01789, \omega_n = 22.36$

The control results are listed in Table 2. It is seen from Tables 1 that vLQG control strategy also is more effective than LQG control strategy for MDOF stochastic vibration system.

Table 2  Numerical results of LQG and vLQG for MDOF: standard deviation

| Displacement (mm) | Without control | LQG | vLQG |
| --- | --- | --- | --- |
| 1st floor | 1.111 | 0.5237 | 0.545 |
| 2nd floor | 2.192 | 0.6834 | 0.6574 |
| 3rd floor | 3.457 | 0.7636 | 0.6837 |
| Velocity (mm/s) | | | |
| 1st floor | 13.56 | 11.51 | 11.67 |
| 2nd floor | 30.80 | 14.93 | 13.60 |
| 3rd floor | 58.21 | 16.61 | 13.59 |
| Control force(kN) | / | 248.3 | 218.2 |

## 5. Conclusion

In the present paper the limitation of classical LQG control strategy for stochastic vibration has been demonstrated.It's control force from displacement is of no effect to depress stochastic response.A modified control strategy which is called vLQG is proposed . A comparison of the numerical results with those by using LQG controller shows that the proposed control strategy vLQG is more effective.